\documentclass[twocolumn,aps,pre,showpacs,floatfix]{revtex4}
\usepackage[latin1]{inputenc}
\usepackage{graphicx}
\usepackage{latexsym}
\usepackage{amssymb}
\usepackage{amsmath}

\def\beq{\begin{eqnarray}}
\def\eeq{\end{eqnarray}}

\def\hx_{\hat\Delta}
\def\hatx_{{\bf \hat x}}
\def\d{{\rm d}}

\def\Tstroke{{\mathcal{T}}}
\def\T{{\boldsymbol{\sf T}}}

\def\bxi{{\boldsymbol\xi}}

\def\B{{\bf B}}
\def\x{{\bf x}}

\def\s{{\bf s}}
\def\f{{\bf F}}

\def\f{{\bf f}}
\def\z{{\bf z}}
\def\A{{\bf A}}

\def\u{{\bf u}}

\def\g{{\bf g}}

\def\flux_{\bar\epsilon}

\def\smal1n{{\scriptscriptstyle (1.1,n)}}

\def\smalze{{\scriptscriptstyle (0)}}
\def\smalun{{\scriptscriptstyle (1)}}

\def\smalCM{{\scriptscriptstyle{\rm CM}}}

 at 24truept
 at 15truept

\begin{document}

\title{Pros and cons of swimming in a noisy environment}
\author{Piero Olla}
\affiliation{ISAC-CNR and INFN, Sez. Cagliari, I--09042 Monserrato, Italy.}
\date{\today}

\begin{abstract}
The problem of optimal microscopic swimming in a noisy environment
is analyzed. A simplified model in which propulsion is generated by the relative motion 
of three spheres connected by immaterial links has been considered. We show
that an optimized noisy microswimmer requires less power for propulsion (on average)
than an optimal noiseless counterpart, migrating with identical mean 
velocity and swimming stroke amplitude. We also show that noise
can be used to overcome some of the limitations of the scallop theorem,
and have a swimmer that is able to propel itself with control over just
one degree of freedom.
\end{abstract}
\maketitle
\section{Introduction}
Microorganisms, such as bacteria and protozoa,
live in a world governed by low-Reynolds number hydrodynamics. 
The strategies for locomotion in such an environment strongly
differ from those valid at macroscopic scales \cite{childress}.  This is exemplified 
by the content of the so called scallop theorem: a sequence of 
deformations in the body of a microswimmer will lead to 
the same displacement, irrespective of the speed at which each deformation 
is carried on. A ``microscopic scallop'' could not propel itself by quickly
closing its valves, and then slowly opening them up to recover its initial
configuration \cite{purcell77}. In order for propulsion to be achieved, 
the microswimmer must carry on a deformation sequence that does not trace 
itself back in time \cite{shapere89}.

Progress in the field of nanotechnology has opened the way to the
possible realization of artificial microswimmers \cite{dreyfus05,yu06,behkam06,leoni09}.
One of the issues that will have to be solved is clearly that of the energy supply. 
This entails the optimization problem of finding the minimum energy strategy
to propel the swimmer at the given velocity. Over the years, much attention has been
given to this problem \cite{lighthill,blake01}.
Most of the effort has been directed to the study of ``deterministic'' microswimmers
\cite{blake01,stone96,becker03,najafi04,avron04}.
If the microswimmer is sufficiently small, however,
thermal fluctuations will start to play a role, and
the optimization problem will turn from deterministic to
stochastic \cite{schweitzer98,vicsek,lobaskin08,golestanian09,dunkel09}. 

This aspect is of relevance to the new field of nanometer-scale swimming, in which
propulsion is achieved by chemical unbalance in the environment, or on the surface
of the nanodevice \cite{golestanian05,paxton06,pooley07}. 
On the other hand, noise is expected to play a role also
at larger scales, as the molecular motors responsible for propulsion, in a microswimmer,
work in and out of equilibrium condition, and are likely to be characterized by fluctuations
of larger amplitude than in thermal equilibrium \cite{ma14}.

The presence of noise will affect the swimmer performance in several ways.
Noise will induce a random component in the swimming 
strokes, and therefore, also in the migration velocity \cite{lobaskin08,dunkel09}.
To this, global diffusion induced by
thermal noise in the fluid, must be added.
An optimal design is likely to require some
minimization of these effects.
At the same time, noise is likely to contribute to the energetics
of the process. Here, things become less clear:
it is well known that there are situations in which 
noise can play a constructive role. Most molecular
motors, indeed, exploit thermal noise in some
way or another to improve their efficiency \cite{kolomeisky07}.

This is precisely the question we want to ask: can
thermal noise be exploited to improve
the swimmer efficiency, and do part of
the job of pushing the device along its desired path? 
The answer is yes, and
we shall see that minimal dissipation, for given 
values of the mean swimming velocity, and of the 
swimming stroke amplitude, is achieved in correspondence 
to a minimum of the swimming velocity fluctuation. This
does not correspond to a minimum of the random component
in the swimming strokes, rather, it is realized through 
optimal control of their
correlations. We shall discuss the nature of 
the internal forces in the swimmer that can 
produce this result.

%
%
To study the problem, we shall consider a swimmer design
that has received much attention recently, namely,
an ensemble of three identical spherical beads
connected by extendable links that do not interact with the fluid \cite{najafi04,earl08}.
Models in this class have been utilized to elucidate 
the properties both of individual swimmers \cite{olla10,friedrich12}
and ensemble of swimmers \cite{alexander08,putz09}, and have found experimental
realization, with optical tweezers used  to drive the beads \cite{leoni09}.

In order to proceed, we shall make a number of simplifying assumption on the structure
of the swimmer, on the stroke amplitude and on the nature of the fluctuations. We shall
assume that the swimmer moving parts and the stroke amplitude are small on the
scale of the swimmer body, and that the dynamics of the system is slow. The first
two assumptions allow a quasilinear description of the dynamics, in which the feedback
of the fluid perturbation on the swimmer dynamics, is disregarded. This corresponds
to the lowest order in the description based on the Kirkwood-Smoluchowski equation,
utilized in \cite{murphy72,dunkel09}. The last assumption of slow dynamics guarantees that the
response of the system to the deformation forces obey standard steady-state 
fluctuation-dissipation relations.


The paper is organized as follows. In Sec. 2, the main results on the optimization of
the deterministic three-bead swimmer are presented.
In Sec. 3, the modification of the problem in the presence
of a noisy component in the swimmer internal dynamics are discussed. In Sec. 4, the
optimization of the noisy swimmer in the weak noise limit is carried on. In Sec. 5,
an example is provided, of how noise can be exploited to simplify the problem of internal control,
by overcoming some of the limitation of the scallop theorem.
Section 6 is devoted to conclusions.

\section{Deterministic case}
\begin{figure}
\begin{center}
\includegraphics[draft=false,width=6.5cm]{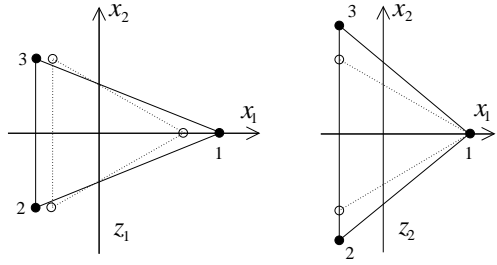}
\caption{Trimer deformations corresponding to $(z_1>0,z_2=0)$ (left) and $(z_1=0,z_2>0)$ (right).
Empty circles indicate the rest shape.
}
\label{noisyfig1}
\end{center}
\end{figure}

The microswimmer design that we are going to consider is a variation on 
the three-sphere model of Najafi \& Golestanian \cite{najafi04,golestanian09a}.
In the original swimmer the three spheres were put on the line, in the present one,
they lie (at rest) at the vertices of an equilateral triangle. A similar model 
was used in \cite{olla10,olla11} to study passive swimming in an external flow.
As in \cite{najafi04}, the beads are supposed identical.
We imagine that the device (called a trimer) 
is constrained to remain with its axis of symmetry along $x_1$,
(see Fig. \ref{noisyfig1}), but that it is otherwise free to translate. Consistent with 
this hypothesis, we assume that the 
trimer can undergo only axisymmetric deformations,
as illustrated in Fig. \ref{noisyfig1}. 
Let us indicate by $\x_i$, $i=1,2,3$, the 
bead coordinates in the comoving frame, and separate the rest component $\x_i^\smalze$:
\beq
\x_1^\smalze&=&R(1/\sqrt{3},0,0), 
\nonumber
\\
\x_2^\smalze&=&R(-1/(2\sqrt{3}),-1/2,0),
\nonumber
\\
\x_3^\smalze&=&R(-1/(2\sqrt{3}),1/2,0),
\label{x0}
\eeq
from the deformation component $\x^\smalun_i=\x_i-\x_i^\smalze$:
\beq
\x_1^\smalun&=&R(z_1/2,0,0), 
\nonumber
\\
\x_2^\smalun&=&(R/4)(-z_1,-\sqrt{3}z_2,0),
\nonumber
\\
\x_3^\smalun&=&(R/4)(-z_1,\sqrt{3}z_2,0).
\label{x1}
\eeq
The parametrization for $\x^\smalun_i$ has been chosen in such a way that dissipation 
is diagonal [see Eq. (\ref{dissipation}) below)].
We shall assume small deformations:
\beq
R^{-1}|\x^\smalun_i|\sim z\ll 1
\nonumber
\eeq
and seek an expression for the migration velocity of the trimer to
lowest order in $z$. 

In creeping flow conditions, forces and particle
velocities are related through the equation
\beq
\dot\x_i=\tilde\u_i(t)+\f_i(t)/\Gamma,
\label{eq2}
\eeq
where $\Gamma$ is the Stokes drag coefficient for the beads and
$\tilde u_i$ is the flow perturbation generated by movement
of the trimer, calculated at $\x_i$ (the fluid is considered quiescent in the absence of
the trimer). For spherical beads, the drag $\Gamma$ can be expressed in terms of the 
solvent kinematic viscosity $\nu_s$ and density $\rho_s$ by means of the formula
$\Gamma=6\pi a\nu_s\rho_s$ \cite{happel}. The flow perturbation
is determined by the instantaneous velocity of the spheres through solution of the Stokes 
equations. 

We assume that the spheres are small compared with the size of the
trimer. This is basically a smallness assumption on the $\tilde\u_i(t)$ 
in Eq. (\ref{eq2}),
that scales indeed with $a/R$, with $a$ the size of the beads.
This allows the forces and the particle velocities to be connected by the linear relation
(summation over repeated indices understood):
\beq
\tilde\u_i(t)&=&\T_{ij}\f_j;
\quad
\T_{ij}\equiv\T(\x_i-\x_j)
\label{utilde}
\\
\T(\x)&
=&\frac{3a}{4\Gamma}\Big[\frac{{\bf 1}}{|\x|}+\frac{\x\x}
{|\x|^3}\Big],
\label{Oseen}
\eeq
where $\T$ is called the Oseen tensor \cite{happel}. 
Notice that, although we are assuming planar deformations,
the equations leading to Eqs. (\ref{utilde}-\ref{Oseen}) are those of 3D 
hydrodynamics.

We define a swimming cycle as 
a closed trajectory in deformation space: 
$\z(t+n\Tstroke)=\z(t)$ $\forall n$,
that results in a finite displacement of the device center of mass
$\x^\smalCM=\frac{1}{3}(\x_1+\x_2+\x_3)$:
$\Delta\x^\smalCM=\x^\smalCM(t+\Tstroke)-\x^\smalCM(t)\ne 0$.
The migration velocity is defined naturally as
$\u^{migr}=\Delta\x^\smalCM/\Tstroke$. In the absence of external forces,
$\sum_i\f_i=0$, we find, from Eq. (\ref{eq2}):
\beq
\u^{migr}=(1/3)\sum_i\langle\tilde\u_i\rangle_{\Tstroke}=\langle\tilde\u_1\rangle_{\Tstroke},
\label{eq3}
\eeq
where $\langle .\rangle_{\Tstroke}$ indicates time average over the stroke time $\Tstroke$.

\begin{figure}
\begin{center}
\includegraphics[draft=false,width=6.5cm]{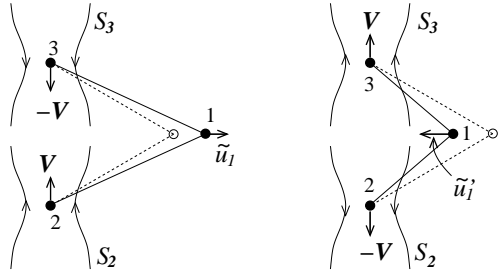}
\caption{Swimming strategy of the microswimmer. The length of 23 is varied 
periodically, $\pi/2$ out of phase with the the two lengths of 12 and 13.  In this
way, when the contraction (extension) speed of $23$ is maximum, the horizontal extension
of the triangle is minimum (maximum).  Field
lines $S_{2,3}$ describe the resulting perturbation in the fluid velocity.  
The asymmetry between the $\tilde u_1$ component experienced in the
two cases causes migration in the negative $x_1$ direction.
}
\label{noisyfig2}
\end{center}
\end{figure}

The deformation sequence responsible for migration, in the case of the trimer, 
is illustrated in Fig. \ref{noisyfig2}.
The flow perturbations $\tilde\u_i$ in Eq. (\ref{eq3}) are expressed in terms of the forces
$\f_i$ by means of the Oseen tensor, Eqs. (\ref{utilde}) and (\ref{Oseen}). 
These in turn can be expressed back
in terms of the deformations, working to lowest order in $a/R$:
\beq
\f_i=\Gamma\dot\x_i=\Gamma\frac{\d\x_i(z_1,z_2)}{\d t}=\s_{ij}\dot z_j,
\label{force}
\eeq
where the constant matrix $\s_{ij}$ is obtained from
Eq. (\ref{x1}).
From Eq. (\ref{utilde}), we obtain the expression, valid to $O(a/R)$:
\beq
\u^{migr}
=\frac{1}{3\Tstroke}\sum_i\oint_\gamma\T_{ij}\s_{jk}\d z_k
\label{eq8}
\eeq
where $\gamma$  indicates the closed path in $z$-space. 

Equation (\ref{eq8}) illustrates
that migration cannot be achieved with a sequence of deformations that traces back in 
time. In particular, this implies that at least two degrees of freedom are necessary
for microscopic swimming (scallop theorem \cite{purcell77,shapere89}). 
In order for the integral along $\gamma$ to give non-zero result, it is necessary that
the integrand is not an exact differential. We can Taylor expand 
the Oseen tensor in $\z$:
\beq
\T_{ij}=\T^\smalze_{ij}+\T^\smalun_{ijk}z_k+\ldots
\nonumber
\eeq
and we see that in order to obtain a non-zero migration velocity, we must keep terms
up to $O(z)$ in the expansion for $\T$:
\beq
u^{migr}&=&\oint A_j\d z_j\equiv
\Phi_{ij}\oint_\gamma z_i\d z_j
\nonumber
\\
\Phi_{ij}&=&\frac{1}{3\Tstroke}\sum_l(\T^\smalun_{lmi}\s_{mj})_1.
\label{potential}
\eeq
From here, simple algebra gives us (see Appendix A):
\beq
u^{migr}=-
\frac{3\sqrt{3}a}{16\Tstroke}\int_0^\Tstroke[z_1\dot z_2-z_2\dot z_1]\d t.
\label{migration}
\eeq
As discussed in \cite{shapere89,avron04}, the expression 
$u^{migr}$
can be interpreted as the flux of
the magnetic field 
\beq
B=\epsilon_{3jk}\partial_{z_j}A_k=-\frac{3\sqrt{3}a}{8}
\eeq
across the surface $S$ in $z$-space, having $\gamma$ for boundary.
Thus, the swimming velocity is purely controlled by the area $S$ and by the time $\Tstroke$
necessary to go through the cycle. 

The work that the trimer must execute to carry out a complete swimming cycle, coincides
with the heat dissipated
\beq
Q=\int_0^{\Tstroke}\f_i(\tau)\cdot\dot\x_i(\tau)\d\tau,
\label{dissipation0}
\eeq
which, from Eqs. (\ref{x1}) and (\ref{force}), can be rewritten in terms of the deformation
\beq
Q=\frac{3\Gamma R^2}{8}\int_0^{\Tstroke}
[(\dot z_1)^2+(\dot z_2)^2]\d\tau.
\label{dissipation}
\eeq
We note at once, by comparison of Eqs. (\ref{migration}) and (\ref{dissipation}), that
the migration velocity and the heat produced in one cycle scale, together with respect
to $z$ and $\Tstroke$: $Q\sim u^{migr}\sim z^2/\Tstroke$, and we have for the dissipated power
$\dot Q\sim Q/\Tstroke$:
\beq
\dot Q\sim (u^{migr})^2/z^2;
\nonumber
\eeq
smaller strokes produce less efficient swimming. Thus, in principle, dissipation could be
sent to zero at finite swimming velocity (forgetting that we are working in a perturbative
regime for $z$), by sending the swimming stroke amplitude and
the stroke time $\Tstroke$ to infinity; a situation resembling the adiabatic ratchet
described in \cite{parrondo98}.

Once the stroke amplitude is fixed, it remains to optimize stroke geometry. From 
$u^{migr}\sim z^2/\Tstroke$, we see that minimizing
expended power at fixed $u^{migr}$ and fixed stroke amplitude, is equivalent to 
minimize the  heat dissipated in a swimming stroke at fixed $u^{migr}$. Formally,
we need to minimize the functional
$\mathcal{A}=Q[\z]+qu^{migr}[\z]$, with respect to $\z$,
with boundary conditions $\z(\Tstroke)=\z(0)$. The constant
$q$ is the Lagrange multiplier required to implement the condition on $u^{migr}$. 

Using Eqs. (\ref{potential}) and (\ref{dissipation}), reabsorbing constants,
and rescaling time in unit of $\Tstroke$, the functional $\mathcal{A}$ takes the
explicit form
\beq
\mathcal{A}=
\int_0^\Tstroke[|\dot\z|^2/2+q\A\cdot\dot\z]\d t.
\label{action}
\eeq
This is the action for a unit mass -- charge $q$ particle, moving in a uniform
magnetic field $\B=\nabla_z\times\A$.
The optimal strategy for the microswimmer
corresponds to the trajectory in deformation space that solves 
the variational problem $\delta\mathcal{A}=0$.
We thus recover the result in \cite{avron04}
that the optimization of the microswimmer can be mapped to the problem
of a charged particle in a uniform magnetic field. 
The fact that $B$ can be approximated as uniform (which is consequence 
of the smallness of $z$), implies that any trajectory $\z(t)$, obtained
from translation of an extremal trajectory for $\mathcal{A}$, will be
extremal as well. The degeneracy can be removed imposing the condition 
that the undeformed state $\z=0$ is really the rest shape for the 
swimmer. This can be expressed as a condition on the time average of the deformation:
\beq
\langle\z\rangle_\Tstroke=0.
\label{degeneracy removal}
\eeq
The end result is uniform circular motion in deformation space, 
with angular frequency $qB$ and center in $\z=0$:
\beq
\z(t)=\bar z_0\Big(\cos(\alpha+qBt),\sin(\alpha+qBt)\Big).
\label{circle}
\eeq
We notice that this solution is characterized by constant dissipated power.
The boundary condition $\z(t+\Tstroke)=\z(t)$ fixes the value of the Lagrange multiplier:
$q=\pm 2\pi/(B\Tstroke)$. 
Using Eq. (\ref{circle}) in Eqs. (\ref{dissipation}) and (\ref{migration}), 
we obtain the optimal values
\beq
\bar Q=
\frac{3\pi^2\Gamma R^2\bar z_0^2}{2\Tstroke},
\quad
\bar u^{migr}=\frac{3\sqrt{3}\pi a\bar z_0^2}{8\Tstroke},
\label{optimized deterministic}
\eeq
where we have taken $q<0$ to have positive $u^{migr}$. Notice that $\bar Q$ and $\bar u^{migr}$
scale together with respect to $z$ and $\Tstroke$, which is a consequence of Eqs. 
(\ref{migration}) and (\ref{dissipation}), i.e. of the small stroke amplitude assumption.
This suggests us to adopt the definition of efficiency given in \cite{shapere87}:
\beq
\eta=u^{migr}/Q,
\label{efficiency}
\eeq
and $\eta$ will be independent of $z$ in the small stroke amplitude regime considered.
Notice that, for fixed $u^{migr}$ and $\Tstroke$, the definition becomes equivalent to the one
in \cite{lighthill}, that was basically $\eta'=(u^{migr})^2/\dot Q$. From Eq.
(\ref{optimized deterministic}), we obtain the optimal efficiency:
\beq
\bar\eta=\bar u^{migr}/\bar Q=\frac{\sqrt{3}}{4\pi}\frac{a}{\Gamma R^2}.
\label{linearQ}
\eeq


\section{Case with thermal noise}
In the presence of noise, quantities entering the optimization procedure, 
such as $Q$ and $u^{migr}$, acquire a fluctuating component. This  forces 
us to work with averages, rather than with instantaneous quantities. [For instance,
Eq. (\ref{degeneracy removal}) will now be understood in a
statistical sense: $\langle\z\rangle=0$. Likewise, the swimming efficiency Eq.
(\ref{efficiency}) will generalize to $\eta=\langle u^{migr}\rangle/\langle Q\rangle$].
At the same time, the swimming stroke will no longer be associated with a closed
orbit in deformation space, and the stroke time $\Tstroke$ must be interpreted
as a characteristic deformation time of the device. 

The choice of the statistical
quantities to minimize, and of the constraints to impose in the optimization 
procedure, strongly rests on the specific problem in which one is interested.
A swimmer that wants to hit a target in the shortest possible time, may have
uncertainty (the variance of $u^{migr}$) among the quantities to minimize or to use
as constraints.
For the long time -- steady state regime we are interested in,
the most natural choice is that of minimizing expended power. We thus impose
minimization of the mean expended power $\langle\dot Q\rangle$ 
at fixed mean migration velocity $\langle u^{migr}\rangle$.
Supported by the observation that optimal noiseless swimming is achieved at 
constant $\dot Q$ [and at constant value of the rate $z_1\dot z_2-z_2\dot z_1$
in Eq. (\ref{migration})], we shall restrict our analysis to a stationary statistics
condition.

We focus on a situation in which the internal forces vary slowly on the scale of the relaxation 
time of the system. This condition appeared to be crucial in stochastic optimization
problems involving a finite time horizon \cite{sivak12},
due to the occurrence of singularities in the solution \cite{schmiedl07}. In 
\cite{aurell12}, the problem was avoided by regularization (see also \cite{aurell11,aurell12a}). 
This is not an issue for the stationary regime considered
here, and the main advantage of an overdamped regime, is that the dynamics is described
by a simple Langevin equation, and that steady state fluctuation-dissipation relations do
apply.

The relaxation time can be estimated, in the case of the trimer, by the 
Stokes time of the beads $\tau_S=m/\Gamma$, where $m$ is the bead mass \cite{happel}.
For spherical beads of radius $a$ and density $\rho_b=\lambda\rho_s$:
$\tau_S=(2/9)\lambda a^2/\nu_s$, with $\nu_s$ and $\rho_s$ the kinematic viscosity
and the density of the solvent. 
Under the condition $\tau_S\ll\Tstroke$, 
the dynamics will be described by a Langevin equation
\beq 
\dot\z+\g=\bxi,
\quad
\langle\xi_i(t)\xi_j(0)\rangle=2K\delta_{ij}\delta(t),
\label{gforce}
\eeq
where, imposing validity of the equilibrium fluctuation-relations, and making use of
Eq. (\ref{dissipation}):
\beq
K=\frac{8k_BT}{3\Gamma R^2},
\label{Ktilde}
\eeq
with $k_B$ the Boltzmann constant and $T$ the temperature \cite{note}.
The mean dissipation is obtained from generalization of Eq. (\ref{dissipation0}):
\beq
\langle\dot Q\rangle=(3\Gamma R^2/8)\langle\dot\z(t)\circ\g(t)\rangle,
\label{heat}
\eeq
where the $\circ$ in the scalar product indicates Stratonovich 
prescription \cite{sekimoto98,schuss}.
Similarly for the mean migration velocity, that is obtained
generalizing Eq. (\ref{migration}):
\beq
\langle u^{migr}\rangle=-
\frac{3\sqrt{3}a}{16}\langle[z_1\dot z_2-z_2\dot z_1]\rangle,
\label{umigr_noise}
\eeq
and it is easy to see, from Eq. (\ref{gforce}), that the term in the average does not 
depend on the choice of the stochastic prescription.

The presence of noise induces a diffusive component in migration, which receives 
contribution both from thermal noise in the fluid and random swimming. The
first contribution to diffusivity can be estimated as
\beq
D^{ext}\sim KR^2.
\label{ext}
\eeq
The second can be obtained from 
$$
D^{int}=\int\d t\langle\hat u(t)\hat u(0)\rangle,
$$
where
$$
\hat u=-\frac{3\sqrt{3}a}{16}[z_1\dot z_2-z_2\dot z_1]-\langle u^{migr}\rangle
$$
is the fluctuating component of the migration velocity.
Working in polar coordinates $(z,\phi)$, we can write
$z_1\dot z_2-z_2\dot z_1=z^2\dot\phi$, and, using Eq. (\ref{gforce}):
\beq
D^{int}&=&
\frac{27a^2}{256}\Big[2K
\nonumber
\\
&+&\int\d t\ \langle z(t)g_\phi(t)z(0)g_\phi(0)\rangle_c\Big].
\label{int}
\eeq
In the present situation, in which noise results from equilibrium
fluctuations in the fluid, and the moving parts in the swimmer are
small, global diffusion, Eq. (\ref{ext}), will dominate over
random swimming, Eq. (\ref{int}). In realistic situations, internal
noise acquires a non-equilibrium component 
\cite{ma14}, and one should
make a substitution $K\to K^{int}\gg K$ in both Eqs. (\ref{gforce}) and
(\ref{int}). (At that point, the possibility of a finite noise correlation time,
should probably be taken into account). Furthermore, an optimal swimmer should have $a\sim R$
[see Eqs. (\ref{optimized deterministic}) and (\ref{umigr_noise})].
Thus, in realistic situations, random swimming is
expected to dominate over global diffusion, $D^{int}\gg D^{ext}$, and
it becomes meaningful to minimize $D^{int}$ along with $\langle Q\rangle$.

\section{Weak noise}
For sufficiently weak noise, we expect that the swimming strategy of the optimal noisy
swimmer, will be sufficiently close to the one in the noiseless case. With this, we intend
that the orbits in deformation space will remain close to the
circular orbit of the optimal noiseless swimmer. 
This imposes the condition on the noise amplitude
\beq
\tilde K=\frac{K}{\omega\bar z_0^2}\ll 1,
\eeq
where $\omega\equiv 2\pi/\Tstroke$ is defined here as the mean circulation frequency
in deformation space; in polar coordinates: 
$\langle\dot\phi\rangle=\omega$.  

\begin{figure}
\begin{center}
\includegraphics[draft=false,width=4cm]{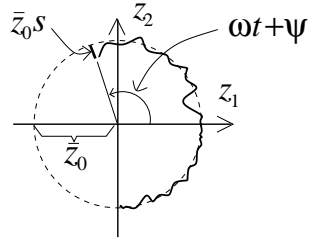}
\caption{Sketch of a $\tilde K\ll 1$ optimal trajectory in deformation space. The phase point
is expected to depart little from the noiseless optimal orbit (the dashed circle). The coordinates
$\psi$ and $\bar z_0s$ give the separation between the phase point of the noisy and the 
deterministic optimal trimer.
}
\label{noisyfig3}
\end{center}
\end{figure}

Working in polar coordinates, the equation of motion (\ref{gforce}) will read:
\beq
&&\dot z=g_z+\frac{K}{z}+\xi_z,
\quad
\dot\phi=\frac{1}{z}g_\phi+\frac{1}{z}\xi_\phi,
\nonumber
\\
&&\langle\xi_i(t)\xi_j(0)\rangle=2K\delta_{ij}\delta(t).
\label{gforce1}
\eeq
In view of a small-noise expansion of Eq. (\ref{gforce1}), we
introduce rescaled variables
$s=(z-\bar z_0)/\bar z_0$, $\psi=\phi-\omega t$, $\tilde t=\omega t$,
giving the deviation of the phase point of the noisy swimmer from its 
optimized noiseless counterpart (see Fig. \ref{noisyfig3}). The forces are
rescaled in consequence:
\beq
\tilde g_s=\frac{g_z}{\bar z_0\omega};
\qquad
\tilde g_\psi=-1+\frac{g_\phi}{z\omega}.
\label{tmp1}
\eeq
Notice that, in
rescaled variables, the definition $\langle\dot\phi\rangle=\omega$ translates in
the constraint $\langle\dot\psi\rangle=0$, i.e. $\langle\tilde g_\psi\rangle=0$.
Consistent with the assumption of small deviation from circular orbit, 
we impose linear dependence of the forces on $s$:
\beq
\tilde g_i=-\alpha_{is}(\psi,\tilde t)s+h_i(\psi,\tilde t),\quad
i=s,\psi,
\label{linear forces}
\eeq
and assume $\alpha_{is}=O(1)$. 

Substituting Eq. (\ref{tmp1}) into Eq. (\ref{gforce1}),
and keeping terms up to $O(\tilde K)$, we 
obtain the equation of motion
\beq
&&\dot s=\tilde g_s+\tilde K+\xi_s;
\quad
\dot\psi=\tilde g_\psi
+\xi_\psi;
\nonumber
\\
&&\langle\xi_i(0)\xi_j(\tilde t)\rangle=2\tilde K\delta_{ij}\delta(\tilde t).
\label{tmp2}
\eeq
The migration velocity is obtained, substituting Eqs. (\ref{gforce1}) and
(\ref{tmp1}) into Eq. (\ref{umigr_noise}):
\beq
\langle u^{migr}\rangle=\bar u^{migr}
\langle (1+s)^2[1+\tilde g_\psi(s,\psi)]\rangle,
\label{tmp0}
\eeq
and we recover, for $\tilde K=0$, Eq. (\ref{optimized deterministic}). 
In analogy with Sec. 2, we choose to minimize $\langle\dot Q\rangle$ at fixed
$\langle z\rangle=\bar z_0$ and $\langle u^{migr}\rangle$, that is
equivalent, from Eqs. (\ref{optimized deterministic}) and (\ref{tmp0}),
to minimizing $\langle Q\rangle=\Tstroke\langle\dot Q\rangle$
at fixed $\langle u^{migr}\rangle$.
Substituting Eqs. (\ref{gforce1}) and
(\ref{tmp1}) into Eq. (\ref{heat}), we obtain:
\beq
\langle Q\rangle&=&\bar Q\Big\langle
\Big[\tilde g^2_s+[(1+s)(1+\tilde g_\psi)]^2
\nonumber
\\
&+&\tilde K\nabla\cdot\tilde\g
\Big]\Big\rangle,
\label{heatheat}
\eeq
where the term $\tilde K\nabla\cdot\tilde\g$ accounts for the correction from the
Stratonovich prescription \cite{schuss}.
Again we recover, for $\tilde K=0$, the expression for dissipation provided in Eq. 
(\ref{optimized deterministic}). 

To determine the difference between minimum dissipation with and without
noise, we fix $\langle u^{migr}\rangle=-\bar u^{migr}$, and study the behavior of
$\mathcal{A}=\langle Q\rangle/\bar Q$,
under the combined constraints $\langle u^{migr}\rangle/\bar u^{migr}=1$ and
$\langle\tilde g_\psi\rangle=0$. Notice that we can write also
$\mathcal{A}=\bar\eta/\eta$,
with $\eta=\langle u^{migr}\rangle/\langle Q\rangle$ and $\bar\eta$ the
optimal efficiency of the deterministic swimmer, given by Eq. (\ref{linearQ}).

Using 
Eqs. (\ref{tmp2}) and (\ref{tmp0}), the two constraints give,
including terms up to $O(\tilde K)$:
\beq
\langle\tilde g_\psi\rangle=0;
\qquad
2\langle s\rangle+\langle s^2\rangle+2\langle s\tilde g_\psi\rangle=0.
\label{constraint}
\eeq
Substituting into Eq. (\ref{heatheat}) and keeping terms up to $O(\tilde K)$, we
finally get:
\beq
\mathcal{A}&=&
\Big\langle\Big\{1+2s\tilde g_\psi+\tilde g^2_s+\tilde g_\psi^2
\nonumber
\\
&+&\tilde K(\partial_s\tilde g_s
+\partial_\psi \tilde g_\psi)\Big\}\Big\rangle,
\label{value function}
\eeq
where the 1 comes from dissipation in the deterministic case.
We see that, contrary to the noiseless case of Eq. (\ref{action}), the forces now enter
explicitly the normalized dissipation $\mathcal{A}$. This reflects the fact that, while in the 
noiseless case one had a single optimal trajectory, in the presence of noise, we
have a distribution of trajectories whose shape is determined by the 
force $\tilde\g$.  
If we restrict the analysis to the case of a stationary swimmer,
the coefficients $\alpha_{is}$ and $h_i$ will be independent of time as well.
We shall consider below two specific driving mechanisms: 
\begin{itemize}
\item
A uniform tangential force pushing the phase point, while some 
constant radial force keeps it close to the unperturbed orbit $z=\bar z_0$.
\item
A potential well confining the phase points both tangentially and radially. The potential
well circulates along the optimal 
noiseless orbit $r=\bar z_0$ with angular frequency $\omega$.
\end{itemize}
In the first case, the stationary distribution of the phase points will be localized in a 
uniform thickness annulus around the optimal noiseless orbit $z=\bar z_0$.  In 
the second case, the stationary distribution will be localized both in $z$ and in $\phi$,
and will circulate with frequency $\omega$ along the orbit $z=\bar z_0$. 

In a realistic swimmer, the driving mechanism may be provided e.g. by a molecular motor
undergoing some cyclic transformation. The two driving mechanisms outlined above, may
correspond therefore to the two regimes of a fluctuating and a fluctuation-free motor respectively.
In the first case, the fluctuations in the motor configuration would sum to those
in the interaction with the swimmer moving parts,
causing global diffusion of $\z$ along the deterministic orbit.
In the second case, the only fluctuation present would be those in the interaction 
between molecular motor and swimmer moving parts, while the molecular motor dynamics
is deterministic.

\subsection{Uniform tangential drive}
In this case, the coefficients $\alpha_{is}$ and $h_i$ in Eq. (\ref{linear forces})
are independent of $\psi$. The linear Langevin equations (\ref{tmp2}) are 
presently solved. At stationary state, the phase points are localized radially,
\beq
\langle s^2\rangle=\frac{\tilde K}{\alpha_{ss}},
\quad
\langle s\rangle=\frac{h_s+\tilde K}{\alpha_{ss}},
\label{tmp31}
\eeq
and uniformly distributed in $\psi$.
Imposing the conditions (\ref{constraint}), we find from Eqs. (\ref{linear forces}) 
and (\ref{tmp2}):
\beq
h_s&=&(\alpha_{\psi s}-3)\tilde K/2,
\nonumber
\\
h_\psi&=&(\alpha_{\psi s}-1)\alpha_{\psi s}\tilde K/(2\alpha_{ss}),
\label{tmp35}
\eeq
and $\alpha_{is}$, $i=s,\psi$, remain 
the only free parameters.
Substituting Eqs. (\ref{tmp31}) and (\ref{tmp35}), together with Eq. 
(\ref{linear forces}), into Eq. (\ref{value function}), and keeping terms up to 
$O(\tilde K)$, we obtain
\beq
\mathcal{A}=1+\frac{\tilde K\alpha_{\psi s}}{\alpha_{ss}}(\alpha_{\psi s}-2).
\label{reduction}
\eeq
For $0<\alpha_{\psi s}<2$, dissipation is reduced with respect to the noiseless case,
the effect being maximum at $\alpha_{\psi s}=1$. In this range, 
efficiency is increased with respect to 
the optimal noiseless case of Eq. (\ref{linearQ}): $\eta>\bar\eta$. 
Dissipation reduction is associated
with decrease of the tangential drive for larger deformations.

\subsection{Circulating potential well}
In this case, the phase points are confined both radially and tangentially, in a potential
well that rotates uniformly with frequency $\omega$. 
If the confinement length is small also tangentially, we can 
linearize the forces also with respect to $\psi$:
\beq
\tilde g_s&=&-\alpha_{ss}(s-\bar s_s)-\alpha_{s\psi}\psi,
\nonumber
\\
\tilde g_\psi&=&-\alpha_{\psi s}(s-\bar s_\psi)-\alpha_{\psi\psi}\psi,
\label{tmp3}
\eeq
where again we assume $\alpha_{ij}=O(1)$.
The equations of motion are still those in Eq. (\ref{tmp2}), and the constraints
in Eq. (\ref{constraint}) continue to apply. As in the case of the
coefficients $h_i$ of Eq. (\ref{tmp35}), it is possible to see that $\bar s_i=O(\tilde K)$,
and therefore also $\langle s\rangle=\langle\psi\rangle=O(\tilde K)$.
Substituting Eq. (\ref{tmp3}) into Eq. (\ref{value function}), we obtain,
keeping terms up to $O(\tilde K)$:
\beq
\mathcal{A}&=&
1+(\alpha_{ss}^2+\alpha_{\psi s}^2-2\alpha_{\psi s})\langle s^2\rangle
\nonumber
\\
&+&(\alpha_{\psi\psi}^2+\alpha_{s\psi}^2)\langle\psi^2\rangle
\nonumber
\\
&+&2[\alpha_{ss}\alpha_{s\psi}
+(\alpha_{\psi s}-1)\alpha_{\psi\psi}]\langle s\psi\rangle
\nonumber
\\
&-&(\alpha_{ss}+\alpha_{\psi\psi})\tilde K.
\label{tmp5}
\eeq
The equation for the correlations entering Eq. (\ref{tmp5})
are obtained from Eqs. (\ref{tmp2}) and (\ref{tmp3}). At stationary state:
\beq
&&\alpha_{ss}\langle s^2\rangle+\alpha_{s\psi}\langle s\psi\rangle=\tilde K,
\nonumber
\\
&&\alpha_{\psi s}\langle s^2\rangle
+(\alpha_{ss}+\alpha_{\psi\psi})\langle s\psi\rangle+
\alpha_{s\psi}\langle\psi^2\rangle=0,
\nonumber
\\
&&\alpha_{\psi s}\langle s\psi\rangle+ \alpha_{\psi\psi}\langle\psi^2\rangle =\tilde K.
\label{tmp4}
\eeq
The solutions of Eq. (\ref{tmp4}) have in general a rather complicated form.
We can solve the system explicitly in some special situation. 

In the case of a purely potential $\tilde\g$, which implies
$\alpha_{s\psi}=\alpha_{\psi s}$, Eq. (\ref{tmp4})
gives
\beq
\langle s^2\rangle&=&
\frac{\alpha_{\psi\psi}\tilde K}{\alpha_{ss}\alpha_{\psi\psi}-\alpha_{s\psi}^2};
\nonumber
\\
\langle s\psi\rangle&=&
\frac{-\alpha_{s\psi}\tilde K}{\alpha_{ss}\alpha_{\psi\psi}-\alpha_{s\psi}^2};
\nonumber
\\
\langle\psi^2\rangle&=&\frac{\alpha_{ss}\tilde K}{\alpha_{ss}\alpha_{\psi\psi}-\alpha_{s\psi}^2}.
\nonumber
\eeq
Substituting into Eq. (\ref{tmp5}), it is possible to see that
$\mathcal{A}=1$, which is the expected result
from a purely potential force.

Things change if we consider a dissipative force.  Let us take for simplicity
$\alpha_{ss}=\alpha_{\psi\psi}$ and $\alpha_{\psi s}=-\alpha_{s\psi}$.  
We obtain
\beq
\langle s^2\rangle=\langle\psi^2\rangle=\tilde K/\alpha_{ss},
\qquad
\langle s\psi\rangle=0,
\nonumber
\eeq
which, upon substitution into Eq. (\ref{tmp5}) gives
\beq
\mathcal{A}=1+2\frac{\alpha_{\psi s}\tilde K}{\alpha_{ss}}(\alpha_{\psi s}-1).
\label{reduction1}
\eeq
Dissipation reduction occurs in this case for $0<\alpha_{\psi s}<1$, corresponding to an
increase of efficiency with respect to the optimal noiseless case $\eta>\bar\eta$.
Maximum reduction occurs at
$\alpha_{\psi s}=1/2$ [compare with Eq. (\ref{reduction})].

\subsection{Randomness minimization}
An optimal microswimmer should have the property of ``arriving at the target on time'', if 
required. Thus, another quantity that one may wish to minimize is the migration velocity 
fluctuation. For simplicity, we keep considering the situation of a stationary swimmer, although the
appropriate setting for such a constraint is that of a device swimming over
a finite distance (or over a finite time interval).

One way to minimize swimming randomness, 
 of course, is to make the deterministic part of the forces, controlling the trimer
deformation, more intense. For fixed strength of the deformation forces, some swimming
strategies will nevertheless lead to a smaller migration velocity fluctuation than others.
We see that in both cases of forcing by a uniform drive, and by a circulating potential well,
minimization of the random migration velocity component is achieved, in good approximation,
together with that of dissipation.

We parametrize the degree of randomness in swimming through the coefficient
$\tilde D=(\omega/(\bar u^{migr})^2)D^{int}$, where $D^{int}$ is defined in Eq. (\ref{int}).
Substituting Eq.
(\ref{tmp1}) into Eq. (\ref{int}), we obtain, keeping terms up to $O(\tilde K)$:
\beq
\tilde D&=&2\tilde K+\int\d\tilde t\ \Big[4\langle s(0)s(\tilde t)\rangle
\nonumber
\\
&+&
4\langle s(0)\tilde g(\tilde t)\rangle+\langle\tilde g_\psi(0)\tilde g_\psi(\tilde t)\rangle\Big].
\label{tilde D}
\eeq
In the uniform tangential drive case, we have $\tilde g_\psi=-\alpha_{\psi s}+O(\tilde K)$.
From Eqs. (\ref{tmp1}) and (\ref{tmp2}), we find 
$\langle s(0)s(\tilde t)\rangle=(\tilde K/\alpha_{ss})\exp(-\alpha_{ss}|\tilde t|)$, and, 
substituting into Eq. (\ref{tilde D}):
\beq
\tilde D=\frac{2\tilde K}{\alpha_{ss}^2}\Big(4+\alpha_{ss}^2
+2\alpha_{\psi s}(\alpha_{\psi s}-2)\Big).
\label{tilde D uniform}
\eeq
Comparing with Eq. (\ref{reduction}), we see that minimal contribution to diffusion from 
random swimming is achieved, together with minimal dissipation, 
for $\alpha_{\psi s}=1$.

In the case of a forcing by a circulating potential well, 
with $\alpha_{\psi\psi}=\alpha_{ss}$ and $\alpha_{s\psi}=\alpha_{\psi s}$, 
we proceed in the same fashion. We have, to lowest order in $\tilde K$:
$\tilde g_s=-\alpha_{ss}s-\alpha_{\psi\psi}\psi$ and
$\tilde g_\psi=-\alpha_{ss}\psi+\alpha_{\psi\psi}s$, which gives, from Eqs.
(\ref{tmp1}) and (\ref{tmp2}):
$\langle s(0)s(\tilde t)\rangle=
\langle \psi(0)\psi(\tilde t)\rangle=
(\tilde K/\alpha_{ss})\exp(-\alpha_{ss}|\tilde t|)$
and $\langle s(0)\psi(\tilde t)\rangle=0$.
Substituting into Eq. (\ref{tilde D}),
we find
\beq
\tilde D=\frac{2\tilde K}{\alpha_{ss}^2}\Big(4+\alpha_{ss}^2
+\alpha_{\psi s}(\alpha_{\psi s}-4)\Big).
\label{tilde D well}
\eeq
Comparing with Eq. (\ref{reduction1}), we find that dissipation reduction implies
random swimming reduction, but not vice versa. In this case, 
minimum random swimming occurs for $\alpha_{\psi s}=2$,
that is out of the domain in which there is dissipation reduction.

\section{Strong noise}
We consider now the situation in which the random component of the deformations
cannot be considered as a perturbation. In \cite{olla11}, it was suggested that noise could
be used to circumvent some of the limitations of the scallop theorem, namely
the need of control over at least two degrees of freedom to achieve locomotion.
We are going to provide an example of this effect, assuming that the only degree 
of freedom acted upon in the trimer by a driving force, is the transversal one $z_2$ (see Fig.
\ref{noisyfig1}), while the longitudinal one $z_1$ is bound by a constant elastic
force. 

We put in Eq. (\ref{gforce}), $g_1=-\omega z_1$ and $g_2=\omega\tilde g(\z)$,
where $\omega$ fixes the deformation time scale of the problem, and $\tilde g(\z)$
contains the drive. The equations of motion thus  become, rescaling time $t\to\tilde t=\omega t$:
\beq
&&\dot z_1+z_1=\xi_1;
\quad
\dot z_2-\tilde g(\z)=\xi_2
\nonumber
\\
&&\langle\xi_i(0)\xi_j(\tilde t)\rangle=2\tilde K\delta_{ij}\delta(\tilde t),
\label{2motor}
\eeq
where now $\tilde K=K/\omega$. 
Proceeding from Eqs. (\ref{heat}) and (\ref{umigr_noise}), we find for the mean expended
power
\beq
\langle\dot Q\rangle=
\frac{3\Gamma\omega^2 R^2}{8}\langle[\tilde g^2+\tilde K\partial_{z_2}\tilde g]\rangle,
\label{Q2}
\eeq
where the $\tilde K\partial_{z_2}\tilde g$ is the correction from the Stratonovich 
prescription \cite{schuss}. Similarly for the mean migration velocity:
\beq
\langle u^{migr}\rangle=-\frac{3\sqrt{3}a\omega}{16}\langle [z_1(z_2+\tilde g)]\rangle.
\label{umigr2}
\eeq
We determine the form of the drive $\tilde g$, 
minimizing $\langle\dot Q\rangle$ at fixed $\langle u^{migr}\rangle$.
The stationary Fokker-Planck equation associated with Eq. (\ref{2motor}) will be
\beq
\mathcal{L}^+\rho=\partial_{z_1}(z_1\rho)-\partial_{z_2}(\tilde g\rho)+\tilde K\nabla_\z^2\rho=0,
\label{Fokker Planck 2}
\eeq
where $\rho=\rho(\z)$ is the stationary probability density function for $\z$.
As we do not know in advance the form of $\rho(\z)$, we must minimize
heat production under the two constraints that $\langle u^{migr}\rangle$ is given, 
and that $\rho$ obeys Eq. (\ref{Fokker Planck 2}). We cannot disregard this last constraint,
as the averages in Eqs. (\ref{Q2}) and (\ref{umigr2}) are carried out precisely with 
$\rho(\z)$. Our cost function will be therefore in the form
\beq
\mathcal{A}&=&\langle\dot Q\rangle-q\langle u^{migr}\rangle
\nonumber
\\
&+&\int\d z_1\d z_2\ J(\z)\mathcal{L}^+\rho(\z),
\nonumber
\eeq
with $J(\z)$ the new Lagrange multiplier, required to guarantee satisfaction locally
of Eq. (\ref{Fokker Planck 2}). Reabsorbing constants in $\mathcal{A}$,
$q$ and $J$, and integrating by parts where necessary, we obtain
\beq
\mathcal{A}
&=&
\int\d z_1\d z_2\ \rho\Big\{\tilde g^2-\tilde K\tilde g\partial_{z_2}\ln\rho
\nonumber
\\
&+&qz_1(z_2+\tilde g)
\nonumber
\\
&+&\Big[\tilde K\nabla_\z^2-z_1\partial_{z_1}+\tilde g\partial_{z_2}\Big]J\Big\}.
\label{cost function 2}
\eeq
Our optimal $g$ is obtained taking the variation of
$\mathcal{A}$ with respect to $g$ and $\rho$, and equating to zero:
\beq
\frac{\delta\mathcal{A}}{\delta \tilde g}&=&\rho\Big[-
2\tilde g+\tilde K\partial_{z_2}\ln\rho-qz_1-\partial_{z_2}J\Big]=0;
\nonumber
\\
\frac{\delta\mathcal{A}}{\delta\rho}&=&
\tilde g^2+\tilde K\partial_{z_2}\tilde g+qz_1(z_2+\tilde g)
\nonumber
\\
&+&\Big[\tilde K\nabla_\z^2-z_1\partial_{z_1}+\tilde g\partial_{z_2}\Big]J=0.
\label{variation}
\eeq
We see that Eq. (\ref{variation}) has solution in the form $\tilde g=-\alpha z_1-\beta z_2$,
with quadratic $\ln\rho$ and $J$. The optimal dynamics is thus linear:
\beq
\dot z_1+z_1=\xi_1;
\quad
\dot z_2+\alpha z_1+\beta z_2=\xi_2.
\label{3motor}
\eeq
The correlation equations associated with Eq. (\ref{3motor}) are
$\langle z_1^2\rangle=\tilde K$;
$(1+\beta)\langle z_1z_2\rangle+\alpha\langle z_1^2\rangle=0$ and
$\alpha\langle z_1z_2\rangle+\beta\langle z_2^2\rangle=\tilde K$, 
that give us:
\beq
&&\langle z_1^2\rangle=\tilde K;
\quad
\langle z_1z_2\rangle=-\frac{\alpha\tilde K}{1+\beta};
\nonumber
\\
&&\langle z_2^2\rangle=\frac{\alpha^2+\beta+1}{\beta(1+\beta)}\tilde K.
\label{correlations}
\eeq
Substituting into Eqs. (\ref{Q2}) and (\ref{umigr2}), we obtain 
\beq
\langle\dot Q\rangle=
\frac{3\Gamma\omega^2R^2}{8}
\frac{\alpha^2\tilde K}{1+\beta},
\label{Q3}
\eeq
and 
\beq
\langle u^{migr}\rangle=-\frac{3\sqrt{3}\omega a}{8}\frac{\alpha\tilde K}{1+\beta}.
\label{umigr3}
\eeq
so that $\langle\dot Q\rangle\sim\alpha\omega\langle u^{migr}\rangle$. 

From Eq. (\ref{3motor}), it appears that $\alpha\omega$ plays the role of circulation 
frequency for the trimer. Its inverse fixes the scale for the stroke time $\Tstroke$. We see that,
as in the deterministic case, expended power can be made smaller by increasing 
$\Tstroke$. For fixed $\langle u^{migr}\rangle$, this will correspond to 
larger swimming strokes. From Eq. (\ref{umigr3}), $\langle u^{migr}\rangle$ fixes in fact
$\alpha/(1+\beta)$, so that  smaller
$\alpha$ will require smaller $\beta$. This in turn corresponds to
larger swimming strokes [see Eq. (\ref{correlations})]. Similarly,
making $\omega$ small, will lead to larger $\tilde K$, and therefore
to larger swimming strokes [see again Eq. (\ref{correlations})]. 

The efficiency of the swimmer $\eta=\langle u^{migr}\rangle/\langle Q\rangle$ can
be determined from Eqs. (\ref{Q3}) and (\ref{umigr3}), once the stroke time $\Tstroke$ is known: 
$\langle Q\rangle=\Tstroke \langle\dot Q\rangle$. Unfortunately, contrary
to the weak noise case, the circulation frequency distribution in deformation space is not
peaked around a well-defined value that could uniquely define the stroke frequency.
(Similarly for the stroke amplitude $z$, that is not peaked around the deterministic
value $\bar z_0$).  We can nevertheless define a stroke time in terms of the 
mean circulation frequency: $\omega\Tstroke=2\pi/\langle\dot\phi\rangle$, $\phi=\tan^{-1}z_2/z_1$,
and compare with the optimal deterministic case.
A calculation, detailed in Appendix B, shows that the efficiency of a swimmer
whose internal dynamics is governed by a linear Langevin equation, such as Eq. (\ref{3motor}),
is always smaller than that of an optimal deterministic swimmer with identical $\Tstroke$. 
This seems to confirm the result in the weak noise regime, that dissipation reduction must
involve some kind of drive reduction at large deformations. This is in fact the opposite of
the situation described in Eq. (\ref{3motor}) (or in any dynamics described by a Langevin
dynamics with center at $\z=0$).


\section{Conclusion}
We have discussed the possibility of dissipation reduction by thermal noise in a simple
microswimmer model. We have shown that an optimal noisy microswimmer will need, for
propulsion at given average swimming velocity and swimming stroke amplitude, 
less energy than its noiseless optimal counterpart, and that the process 
goes together with reduction in randomness of the swimming velocity.
Another effect of noise is that 
some of the constraints of the scallop theorem can be bypassed, as the 
swimmer can propel itself with control over just one degree of freedom.

The optimal design of a noisy microswimmer, imposes
constraints on the functional form of the deformation forces driving its dynamics,
that are not present in the deterministic case \cite{avron04}. The need
to optimize a distribution of deformation sequences, rather than a single deformation
sequence stands at the basis of the new constraints. We have determined the optimal 
deformation force profiles in both strong and weak noise conditions.

We stress that the results obtained are valid only for
a regime of small moving parts and small swimming strokes.
A detailed numerical analysis would be required to confirm our results, in the
case of an optimal microswimmer with moving parts and swimming
strokes comparable in size with the swimmer body. 

Another issue that should probably be addressed is the robustness of the results,
with respect to modification in the form of the noise
[e.g. as regards possible finiteness of the correlation time in Eq. (\ref{gforce})].

Throughout this paper, we have considered the case of a stationary swimmer, which 
is consistent with focus on average quantities such as the mean 
expended power $\langle\dot Q\rangle$ and the mean migration velocity
$\langle u^{migr}\rangle$. 
In a general finite time horizon situation, the statistics
will be time dependent, and other quantities beyond $\langle\dot Q\rangle$  and
$\langle u^{migr}\rangle$ are expected to play a role in the optimization process.

We must remember that, even in the case of an infinite time horizon,  stationarity
is an assumption. We have not examined e.g. the case of a swimmer for which
$\langle\dot Q\rangle$ and $\langle u^{migr}\rangle$ have a component that
is periodic in time. Minimization in this case would be still performed
working with the constant components of $\langle\dot Q\rangle$ and $\langle u^{migr}\rangle$, 
but the space of possible swimming
strategies is larger than in the constant case.
Thus, we do not rule out the possibility
that swimming strategies admitting periodic components in $\langle\dot Q\rangle$ and 
$\langle u^{migr}\rangle$ may have better properties than the ones
considered in the present analysis.

\appendix

\section{Determination of the migration velocity in the deterministic case}
The migration velocity in the deterministic case
is obtained substituting Eqs. (\ref{eq2}) and (\ref{utilde}) into Eq. 
(\ref{eq3}). Exploiting
equality of the contribution to $\tilde\u_1$ from particles 2 and 3, we can write:
\beq
u^{migr}\simeq 2\Gamma
\langle[T^\smalun_{12,11}\dot x^\smalun_{2,1}
+T^\smalun_{12,12}\dot x_{2,2}^\smalun]\rangle_\Tstroke,
\label{u^migr tmp}
\eeq
and we adopt the convention that, 
in vector and tensor expressions, indices before comma indicate particle labels; after comma,
they indicate vector components.

A little algebra from Eqs. (\ref{x0}-\ref{x1}) and (\ref{utilde}-\ref{Oseen}) gives us
\beq
T_{12,11}&=&\sigma\Big(\frac{7}{4}-\frac{\sqrt{3}}{32}(9z_1-5z_2)\Big),
\nonumber
\\
T_{12,12}&=&\sigma\Big(\frac{3\sqrt{3}}{16}-\frac{9}{32}(3z_1+z_2)\Big),
\label{Oseen explicit}
\eeq
where $\sigma=3a/(4\Gamma R)$.
Substituting Eq. (\ref{Oseen explicit}) into Eq. (\ref{u^migr tmp}), we obtain 
\beq
u^{migr}&\simeq&\frac{3\sqrt{3}a}{265}\langle 5z_2\dot z_1-27z_1\dot z_2\rangle_\Tstroke
\nonumber
\\
&=&-\frac{3\sqrt{3}a}{16}\langle z_1\dot z_2-z_2\dot z_1\rangle_\Tstroke,
\label{lelle}
\eeq
that is Eq. (\ref{migration}).

\section{Swimmer efficiency in the strong noise regime}
\begin{figure}
\begin{center}
\includegraphics[draft=false,width=6.5cm]{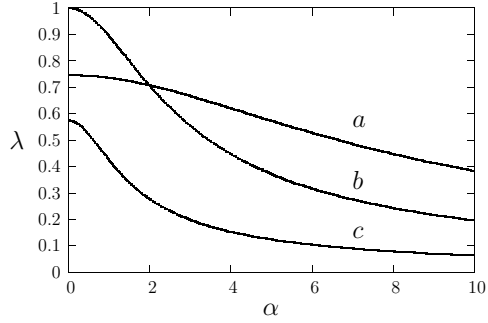}
\caption{Plot of the ratio $\lambda=\eta/\bar\eta$ vs $\alpha$ for three different values
of $\beta$: ($a$) $\beta=5.0$; ($b$) $\beta=1.0$; ($c$) $\beta=0.1$.
}
\label{noisyfig4}
\end{center}
\end{figure}

The mean circulation frequency can be obtained from Eq. (\ref{3motor}):
\beq 
\langle\dot\phi\rangle
=(1-\beta)\langle\sin\phi\cos\phi\rangle-\alpha\langle(\cos\phi)^2\rangle.
\label{B1}
\eeq
The angular distribution $\rho(\phi)$ to be used in Eq. (\ref{B1}), 
is obtained from the deformation PDF $\rho(\z)$:
$\rho(\phi)=\int_0^\infty z\d z\ \rho(\z)$;
$\rho(\z)=A\exp(-(1/2)Z^{-1}_{ij}z_iz_j)$, $Z_{ij}=\langle z_iz_j\rangle$.
Using Eq. (\ref{correlations}), we obtain, after little algebra:
\beq
\rho(\phi)&=&B[C(\cos\phi)^2
\nonumber
\\
&+&D(\sin\phi)^2+E\sin\phi\cos\phi]^{-1},
\label{B2}
\eeq
where
\beq
&&C=1+\beta+\alpha^2;\quad
D=(1+\beta)\beta;
\nonumber
\\
&&E=2\alpha\beta;
\label{B3}
\eeq
and $B$ is a normalization.
Adopting as definition of the stroke time: $\omega\Tstroke=2\pi/\langle\dot\phi\rangle$, 
Eqs. (\ref{linearQ}), (\ref{Q3}) and (\ref{umigr3}), give
us, for the efficiency in the strong noise regime:
\beq
\lambda=\eta/\bar\eta=2\langle\dot\phi\rangle/\alpha.
\label{B4}
\eeq
Evaluation of the average in Eq. (\ref{B1}), with the distribution in Eqs. (\ref{B2}) and
(\ref{B3}), gives the result in Fig. \ref{noisyfig4}.
Equation (\ref{3motor}) leads to an efficiency that is always below that of the corresponding
optimal deterministic swimmer. Maximum efficiency is achieved
for $\beta=1$ and $\alpha$ small, which is, in some sense, a 
maximally isotropic forcing in deformation space. This is not a casual result.
It is possible to see that a trimer
obeying a symmetrized version of Eq. (\ref{3motor}): 
$\dot z_1+z_1-\alpha z_1=\xi_1$, 
$\dot z_2+z_1+\alpha z_2=\xi_2$, 
is characterized by efficiency $\eta=\bar\eta$
for all values of $\alpha$.

\acknowledgements
I wish to thank Paolo Muratore Ginanneschi and Carlos Mejia Monasterio for
interesting and helpful conversation. This research was carried on in part
at the Mathematics Department of the University of Helsinki, with 
financial support by the Center 
of Excellence ``Analysis and Dynamics'' of the Academy of Finland.


\begin{thebibliography}{10}
\bibitem{note} It is possible to see, from Eq. (\ref{x1}), that the
kinetic energy of the trimer is diagonal in the center of mass velocity
$\dot x_1^\smalCM$ and the deformation rates $\dot z_{1,2}$:
$E=(3/2)m(\dot x_1^\smalCM)^2+(3/8)mR^2(\dot z_1^2+\dot z_2^2)$ (to lowest order in $a/R$, we
disregard the contribution from $\tilde\u_i$ to the bead velocities). The standard argument
that each degree of freedom contributes $k_BT/2$ to the energy fluctuations then applies,
and leads to Eqs. (\ref{gforce}) and (\ref{Ktilde}).
\end{thebibliography}
\end{document}